\documentclass[12pt]{article}
\usepackage{microtype} \DisableLigatures{encoding = *, family = * }
\usepackage{subeqn}
\usepackage{graphicx}
\usepackage{pstricks}
\usepackage{amsfonts,amssymb,slashbox}
\usepackage{mathptmx,helvet,courier,makeidx,multicol,footmisc}
\usepackage[numbers]{natbib}
\bibpunct{(}{)}{;}{a}{,}{,}

\newcommand{\commentout}[1]{}

\newcommand{\ba}{\begin{array}}
        \newcommand{\ea}{\end{array}}
\newcommand{\bc}{\begin{center}}
        \newcommand{\ec}{\end{center}}
\newcommand{\bdm}{\begin{displaymath}}
        \newcommand{\edm}{\end{displaymath}}
\newcommand{\bds} {\begin{description}}
        \newcommand{\eds} {\end{description}}
\newcommand{\ben}{\begin{enumerate}}
        \newcommand{\een}{\end{enumerate}}
\newcommand{\beq}{\begin{equation}}
        \newcommand{\eeq}{\end{equation}}
\newcommand{\bfg} {\begin{figure}[fh]}
        \newcommand{\efg} {\end{figure}}
\newcommand{\bi} {\begin {itemize}}
        \newcommand{\ei} {\end {itemize}}
\newcommand{\bqn}{\begin{eqnarray}}
        \newcommand{\eqn}{\end{eqnarray}}
\newcommand{\bqs}{\begin{eqnarray*}}
        \newcommand{\eqs}{\end{eqnarray*}}
\newcommand{\bsl} {\begin{slide}[8.8in,6.7in]}
        \newcommand{\esl} {\end{slide}}
\newcommand{\bss} {\begin{slide*}[9.3in,6.7in]}
        \newcommand{\ess} {\end{slide*}}
\newcommand{\bsq}{\begin{subequations}}
        \newcommand{\esq}{\end{subequations}}       
\newcommand{\btb} {\begin {table}}
        \newcommand{\etb} {\end {table}}
\newcommand{\bvb}{\begin{verbatim}}
        \newcommand{\evb}{\end{verbatim}}

\newcommand{\m}{\mbox}
\newcommand {\der}[2] {{\frac {\m {d} {#1}} {\m{d} {#2}}}}

\newcommand {\pd}[2] {{\frac {\partial {#1}} {\partial {#2}}}}

\newcommand{\cas}[1]{{{\left \{ \ba #1 \ea \right. }}}

\newcommand{\reff}[1] {{{Figure \ref {#1}}}}
\newcommand{\refe}[1] {{(\ref {#1})}}

\def\pmb#1{\setbox0=\hbox{$#1$}%
   \kern-.025em\copy0\kern-\wd0
   \kern.05em\copy0\kern-\wd0
   \kern-.025em\raise.0433em\box0 }

\def\eop{{\hfill $\blacksquare$}}
\def\r{{\rho}}
\newtheorem{theorem}{Theorem}[section]

\def\dx     {{\Delta x}}
\def\dt     {{\Delta t}}

\begin{document}
\title{A kinematic wave theory of capacity drop} 
\author{Wen-Long Jin \footnote{Department of Civil and Environmental Engineering, California Institute for Telecommunications and Information Technology, Institute of Transportation Studies, 4000 Anteater Instruction and Research Bldg, University of California, Irvine, CA 92697-3600. Tel: 949-824-1672. Fax: 949-824-8385. Email: wjin@uci.edu. Corresponding author} and Qi-Jian Gan \footnote{Department of Civil and Environmental Engineering, Institute of Transportation Studies, University of California, Irvine, CA 92697-3600. Email: qgan@uci.edu} and Jean-Patrick Lebacque \footnote{Universit\'e Paris-Est, IFSTTAR, GRETTIA, 14-20 Boulevard Newton, Cit\'e Descartes, Champs sur Marne, 77447 Marne la Vall\'ee Cedex 2, France. Email: jean-patrick.lebacque@ifsttar.fr}}
\maketitle
\begin{abstract}
Capacity drop at active bottlenecks is one of the most puzzling traffic phenomena, but a thorough understanding is practically important for designing variable speed limit and ramp metering strategies. In this study, we attempt to develop a simple model of capacity drop within the framework of kinematic wave theory based on the observation that capacity drop occurs when an upstream queue forms at an active bottleneck. In addition, we assume that the fundamental diagrams are continuous in steady states. This assumption is consistent with observations and can avoid unrealistic infinite characteristic wave speeds in discontinuous fundamental diagrams.  
A core component of the new model is an entropy condition defined by a discontinuous boundary flux function. For a lane-drop area, we demonstrate that the model is well-defined, and its Riemann problem can be uniquely solved. 
We theoretically discuss traffic stability with this model subject to perturbations in density, upstream demand, and downstream supply. We clarify that discontinuous flow-density relations, or so-called ``discontinuous'' fundamental diagrams, are caused by incomplete observations of traffic states. Theoretical results are consistent with observations in the literature and are verified by numerical simulations and empirical observations. We finally discuss potential applications and future studies.

\end{abstract}

\section{Introduction}
Since the 1990s, the so-called two-capacity or capacity-drop phenomenon of active bottlenecks, in which ``maximum flow rates decrease when queues form'', has been observed for many decades and verified at many merge locations  \citep{banks1990flow,banks1991twocapacity,hall1991capacity}.
That is, when the total demand of the upstream mainline freeway and the on-ramp exceeds the capacity of the downstream mainline freeway, a queue forms on the mainline freeway, and the discharging flow-rate drops below the capacity of the downstream mainline freeway.  
Such ``capacity drop'' has been observed at merges, tunnels, lane drops, curves, and upgrades, where the bottlenecks cannot provide sufficient space for upstream vehicles \citep{chung2007relation}.  Capacity drop also occurs at bottlenecks caused by work zones \citep{krammes1994updated,dixon1996capacity,jiang1999traffic} as well as accidents/incidents \citep{smith2003characterization}.
 
\bfg\bc
\includegraphics[width=3in]{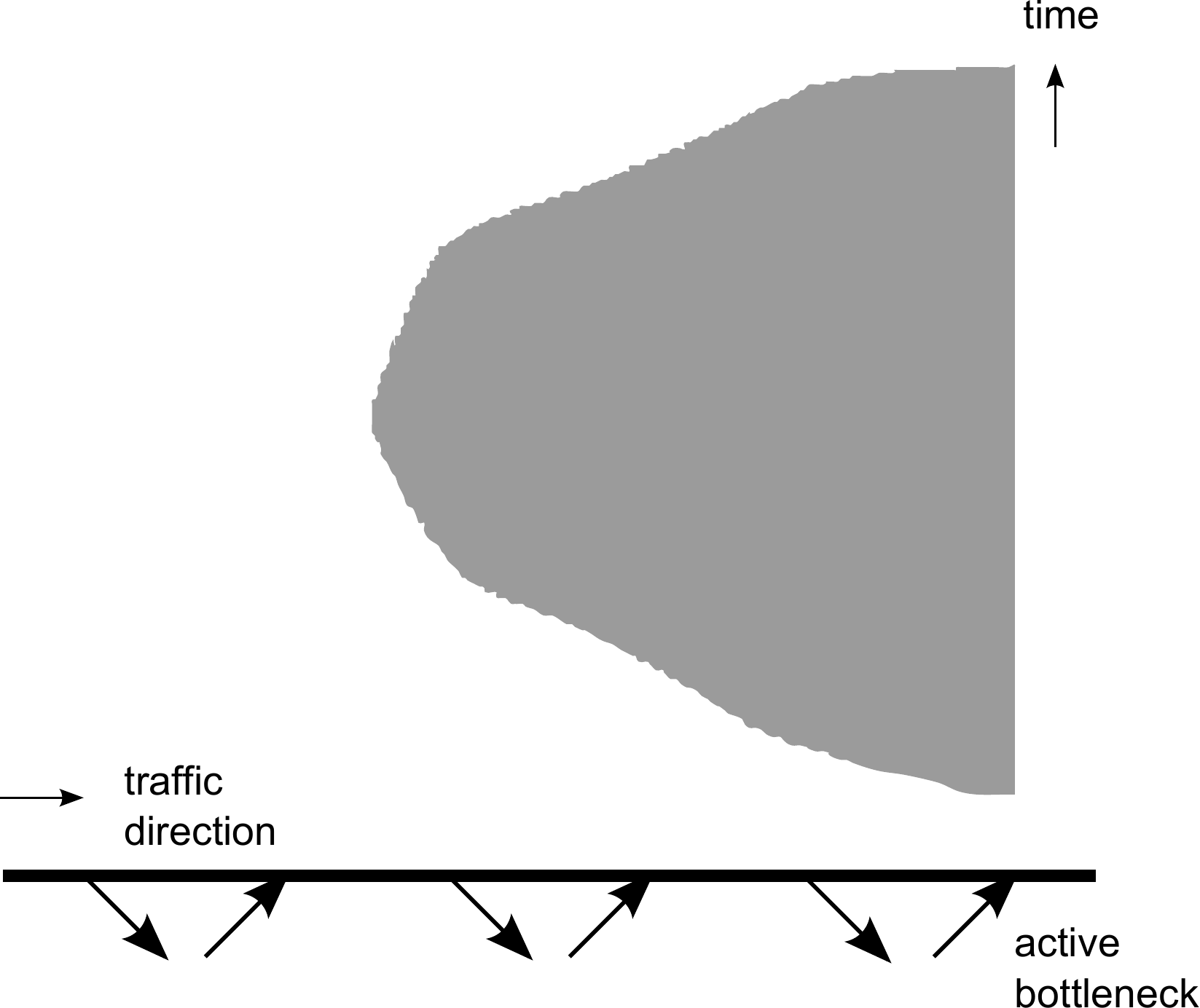}\caption{Traffic queue (the gray region) at an active bottleneck}\label{active_bottleneck}
\ec\efg

 As shown in \reff{active_bottleneck}, a drop at the downstream bottleneck's discharging flow-rate can reduce the total discharge rate of the whole corridor, including at impacted off-ramps, and prolong vehicles' travel times \citep{daganzo1999remarks}.
 That the capacity of a road network may drop substantially when it is most needed during the peak period has been a baffling nature of freeway traffic dynamics \citep{papageorgiou2002freeway}. 
Hence an important motivation and theoretical foundation for developing ramp metering, variable speed limits, and other control strategies is to avoid or delay the occurrence of capacity drop \citep{banks1991metering,papageorgiou1991alinea,papageorgiou1997ALINEA,cassidy2005merge,papageorgiou2005review,papageorgiou2007alinea}.

Since 1960s, it has been observed that the flow-density relation can be discontinuous in a reverse-lambda shape \citep{edie1961cf,drake1967fd,koshi1983fd,payne1984discontinuity,hall1992fd}.
In \citep{hall1991capacity}, it was shown that such discontinuous fundamental diagrams generally arise in the congested area of an active bottleneck and suggested that the discontinuity is caused by the capacity drop phenomenon. 
Therefore, many models of capacity drop have been based on the assumption of discontinuous fundamental diagrams. For example, in \citep{lu2008explicit,lu2009entropy}, an attempt was made to describe capacity drop with discontinuous fundamental diagrams within the framework of the LWR model \citep{lighthill1955lwr,richards1956lwr}. 

However, a discontinuous fundamental diagram is challenged both theoretically and empirically. Theoretically, a discontinuous flow-density relation is non-differentiable at the discontinuous point and leads to infinite characteristic wave speeds \citep{li2013modeling}. Empirically,  \citet{cassidy1998bivariate} demonstrated that bivariate fundamental diagrams are still continuous if one excludes non-stationary data, even near a bottleneck with capacity drop, and this suggests that the flow-density relation is not discontinuous, but capacity drop prevents the occurrence of some intermediate traffic states.

\begin{figure}\bc
\includegraphics[width=3in]{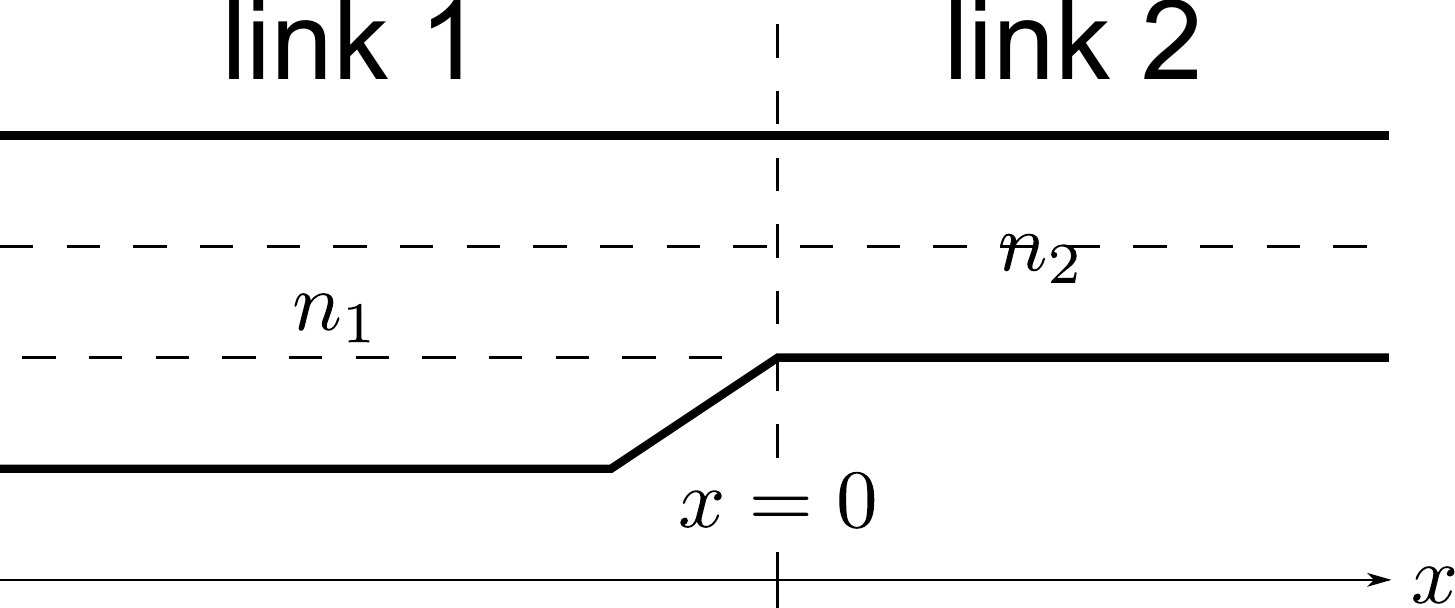}\caption{A lane-drop bottleneck} \label{lane-drop}
\ec
\end{figure}

In this study we propose a new model of capacity drop to reconcile the discontinuous fundamental diagrams with capacity drop. For an active lane-drop bottleneck, as shown in \reff{lane-drop}, we attempt to develop a simple model of capacity drop within the framework of kinematic wave theory developed in \citep{jin2009sd,jin2012network}, in which the junction flux function in term of upstream demands and downstream supplies is used as an entropy condition to pick out unique, physical solutions. Here we introduce a new flux function based on the observation that upstream congestion and capacity drop occur immediately after the upstream demand exceeds the downstream supply. 

Different from existing models of capacity drop \citep{lu2008explicit,lu2009entropy}, the new model still uses continuous fundamental diagrams for the upstream and downstream links. 
However, the new flux function is a discontinuous function in upstream demand and downstream supply. This is different from traditional flux functions, which are generally continuous  \citep{daganzo1995ctm,lebacque1996godunov,jin2003merge,ni2005merge,lebacque2005network,jin2010merge,tampere2011generic,jin2012network}.
Therefore the model is capable of reproducing the characteristics of the capacity drop phenomenon: (i) capacity drop at active bottlenecks occurs when the upstream traffic is congested; (ii) when congestion arises, the maximum discharging flow-rate cannot exceed a dropped capacity; and (iii) the observed flow-density relation is discontinuous.
In addition, the new capacity drop model can be readily incorporated in the Cell Transmission Model (CTM) to simulate impacts of capacity drop on traffic dynamics \citep{daganzo1995ctm}. 

In the literature, there have been many studies on the mechanism of capacity drop. It was observed that, when an active bottleneck stabilizes, there is a long gradually accelerating region around the bottleneck \citep{banks1991twocapacity}, and it was conjectured that the reduced flow is a consequence of the way drivers
accelerate away from the queue \citep{hall1991capacity,papageorgiou2008real}. In \citep{cassidy2005merge}, it was observed that the occurrence of capacity drop at a merging bottleneck is associated with an extensive queue on the shoulder lane upstream to the merging point, sharp declines in vehicle speeds, and increases in lane-changing activities. However, it was pointed out that lane changing alone might not explain the capacity drop. Even though there have been many studies on capacity drop caused by heterogeneous drivers \citep{daganzo2002behavior,chung2004test}, pedestrians \citep{jiang2002capacity}, buses \citep{zhao2007capacity}, or accidents \citep{knoop2008capacity}, the causes and mechanism of capacity drop at active bottlenecks remain to be revealed.  In \citep{persaud1998exploration,persaud2001breakdown}, traffic breakdown and capacity drop were found to be related to the upstream traffic demand randomly. The phenomenon has been successfully replicated in microscopic or hybrid simulations \citep{tampere2005behavioural,treiber2006understanding,laval2006lc,carlson2010optimal,leclercq2011capacity}.
In contrast, in this study we focus on replicating the phenomenological characteristics of capacity drop.

Note that, in \citep{carlson2010optimal,parzani2012second}, higher-order continuum models were shown to replicate capacity drop, but the capacity in higher-order models may be different from the generally used value in steady states \citep{zhang2001note}. Theoretically, the capacity of a road section should be the maximum flow-rate that can be achieved in stationary traffic along an infinitely long, homogeneous road with the same parameters, including the number of lanes, speed limit, grade, curvature, and so on. In addition, the new model is still first-order and, therefore, more efficient computationally and tractable mathematically.
 
The rest of the paper is organized as follows. In Section 2, we present a new model for capacity drop at a lane-drop bottleneck and demonstrate that it is well-defined. In Section 3, we further discuss the analytical properties of the new model. In Section 4, we study traffic statics and dynamics on a ring road with capacity drop. In Section 5, we present an empirical study to validate the new model. In Section 6, we make some concluding remarks. 

\section{A kinematic wave model of capacity drop at a lane-drop bottleneck}
For a road with a lane-drop bottleneck, shown in \reff{lane-drop}, the upstream link 1 has $n_1$ lanes and the downstream link 2 has $n_2<n_1$ lanes. Here we omit the dynamics as well as impacts of the transition region from $n_1$ lanes to $n_2$ lanes and assume that the lane-drop bottleneck is at $x=0$. 

We denote traffic density, speed, and flow-rate by $k(x,t)$, $v(x,t)$, $q(x,t)$ respectively, which are all functions of location $x$ and time $t$. The number of lanes at $x$ is denoted by $n(x)$. Hereafter we will omit $(x,t)$ from these variables unless necessary. Then the LWR model of traffic flow on the road shown in \reff{lane-drop} can be defined by the following rules:
\bi
\item [R1.] The constitutive law of continuous media: $q=kv$.
\item [R2.] The location-dependent fundamental diagram \citep{greenshields1935capacity}: $v=V(n,k)$ and $q=k V(n,k)=Q(n,k)$.
\item [R3.] The continuity equation: $\pd {k}t+\pd{q}x=0$.
\item [R4.] The existence of weak solutions: discontinuous shock waves can develop from continuous initial conditions.
\item [R5.] The entropy condition: unique, physical solutions of the LWR model should satisfy an entropy condition. 
\ei
The first three rules lead to the following inhomogeneous LWR model
\bqn
\pd{k}t+\pd{Q(n,k)}x&=&0, \label{inhlwr}
\eqn 
which is a hyperbolic conservation law. 
Among the five rules, R1,R3, and R4 are generic for all continuum dynamics, but R2 and R5 are system dependent. For a traffic system, R2 describes steady characteristics, in terms of flow- and speed-density relations, and R5 describes dynamic car-following, lane-changing, merging, diverging, and other driving behaviors. 

Traditionally, the fundamental diagram is continuous \citep{delcastillo1995fd_empirical}. However, there have been many evidences of discontinuous fundamental diagrams \citep{edie1961cf,drake1967fd,koshi1983fd,payne1984discontinuity,hall1992fd}, and it appears that capacity drop is one of the reasons to cause such discontinuous fundamental diagrams \citep{hall1991capacity}. Even though empirically appealing, theoretically such discontinuous fundamental diagrams lead to infinite characteristic speeds at the discontinuous point \citep{li2013modeling}, and empirically it appears that the steady relation can still be continuous even when capacity drop occurs \citep{cassidy1998bivariate}. Therefore, it would be appealing if we can explain the capacity drop phenomenon without such a discontinuous fundamental diagram.

For the  inhomogeneous LWR model \refe{inhlwr}, traditional entropy conditions based on characteristics were discussed in \citep{jin2003inhlwr}. In \citep{newell1993sim,daganzo2006variational}, a variational principle was proposed to uniquely solve \refe{inhlwr}. In \citep{jin2009sd}, it was shown that the boundary flux function, which was initially introduced in CTM \citep{daganzo1995ctm}, can be used as an entropy condition. 
However, a reasonable entropy condition to capture capacity drop was not discussed in the aforementioned studies.

In this study, we attempt to model capacity drop by applying traditional continuous fundamental diagrams for both the upstream and downstream links at a lane-drop bottleneck, but introducing a discontinuous entropy condition in terms of a boundary flux function in upstream demands and downstream supplies. Apparently such a continuous fundamental diagram is devoid of infinite characteristic speeds, and we will demonstrate that the discontinuous flux function is capable of capturing the major features of capacity drop: capacity drop occurs with upstream congestion. But this model is phenomenological since (i) the capacity drop magnitude is exogenous, and the capacity drop occurs immediately following the upstream congestion and exactly at the lane-drop location, $x=0$.

\subsection{A discontinuous entropy condition}

We denote traffic demand and supply at $(x,t)$ by $d(x,t)$ and $s(x,t)$, respectively.
For a continuous flow-density relation $Q(n,k)$, which is unimodal in $k$, traffic demand and supply are respectively its increasing and decreasing branches \citep{engquist1980calculation,daganzo1995ctm,lebacque1996godunov}:
\bqs
d&=&D(n,k)\equiv Q(n,\min\{k_c(n),k\}),\\
s&=&S(n,k)\equiv Q(n,\max\{k_c(n),k\}),
\eqs
where $k_c(n)$ is the critical density for $n$ lanes.
An example is the following triangular fundamental diagram, which has been derived from car-following models and verified by observations \citep{munjal1971multilane,haberman1977model,newell1993sim}:
\bqn
Q(n,k)&=&\min\left\{v^* k, \frac 1\tau (n-\frac{k}{k^*})\right\}, \label{tri-fd}
\eqn 
where $v^*$ is the free-flow speed, $\tau$ the time-gap, $k^*$ the jam density per lane, and $k_c(n)=\frac{n k^*}{1+\tau v^* k^*}$.
Since $Q(n,k)$ is unimodal in $k$, $D(n,k)/S(n,k)$ is a strictly increasing function in $k$. If we define the congestion level by $\gamma=d/s$, then traffic density is a function of $\gamma$
\bqn
k&=&K(n,\gamma),
\eqn
such that $D(n,k)/S(n,k)=\gamma$.

Based on the definitions of traffic demand and supply, in \citep{jin2009sd} it was shown that the following flux function is a valid entropy condition for the inhomogeneous LWR model \refe{inhlwr}:
\bqn
q(x,t)&=&\min\{d(x^-,t),s(x^+,t)\}, \label{ctm-entropy}
\eqn 
where $d(x^-,t)$ and $s(x^+,t)$ are the upstream demand and downstream supply at $x$.  
That is, the LWR model, \refe{inhlwr}, coupled with \refe{ctm-entropy} has unique weak solutions with given initial and boundary conditions \citep{holden1995unidirection}. In addition, \refe{ctm-entropy} is consistent with the traditional entropy conditions by \citep{lax1972shock}, \citep{ansorge1990entropy}, and  \citep{isaacson1992resonance}. 
We can see that, when the downstream link is uncongested, the maximum throughput of the lane-drop bottleneck is the capacity of the downstream link $C_2=Q(n_2,k_c(n_2))$. Therefore \refe{ctm-entropy} cannot model the capacity drop phenomenon. 
In \citep{jin2013multi}, it was shown that systematic lane changes can reduce $C_2$, which can be computed from the number of lanes, $n_1$ and $n_2$, the average duration of each lane change, and the length of the lane-changing region. However, since \refe{ctm-entropy} can still be applied to model capacity reduction caused by lane changes, the phenomenon of capacity drop was not captured.

Since capacity drop arises with a queue on the upstream link 1, it is associated with the traffic dynamics at the transition region at $x=0$ between the two links, and it is reasonable to modify the entropy condition, \refe{ctm-entropy}, to capture this dynamic feature. Here we still apply \refe{ctm-entropy} as the entropy condition for 
traffic inside the upstream link 1 and downstream link 2, but introduce the following new entropy condition for the transition region at $x=0$:
\bqn
q(0,t)&=&\cas{{ll} d(0^-,t), & d(0^-,t)\leq s(0^+,t) \\ \min\{s(0^+,t),C_*\}, & d(0^-,t)>s(0^+,t)} \label{cd-entropy}
\eqn
where $d(0^-,t)$ is the upstream demand, $s(0^+,t)$ the downstream supply, and $C_*$ the dropped capacity. Here we assume that $C_*<C_2$, and the capacity-drop ratio is defined by
\bqs
\Delta&=&1-\frac{C_*}{C_2}.
\eqs
Based on the observation that the maximum flow-rate for the bottlenecks can reach 2300 vphpl in free-flow traffic  \citep{hcm1985,hall1991capacity}, capacity drop magnitudes have been quantified for different locations.
 Generally, the magnitude of capacity drop is in the order of 10\%, even 20\% \citep{persaud1998exploration,cassidy1999bottlenecks, bertini2005empirical,chung2007relation}, and such a drop is stable, although interactions among several bottlenecks can cause fluctuations in discharging flow-rates \citep{kim2012capacity}.

If we introduce an indicator function, $I_{d(0^-,t)>s(0^+,t)}$, which equals 1 if $d(0^-,t)>s(0^+,t)$ and 0 otherwise, then \refe{cd-entropy} can be re-written as
\bqs
q(0,t)&=&\min\{d(0^-,t), s(0^+,t), C_2(1-\Delta \cdot I_{d(0^-,t)>s(0^+,t)})\}.
\eqs
We can see that the new flux function, i.e., entropy condition, is consistent with the following macroscopic rules:
\ben
\item The flux is maximized: $\max q(0,t)$.
\item The flux is not greater than the upstream demand or the downstream supply: $q(0,t)\leq d(0^-,t)$, and $q(0,t)\leq s(0^+,t)$.
\item When the upstream link is congested, the flux is not greater than the dropped capacity: $q(0,t)\leq C_2(1-\Delta \cdot I_{d(0^-,t)>s(0^+,t)})$.
\een
Therefore the new entropy condition is equivalent to the following optimization problem:
\bqn
\max q(0,t), \label{cd-entropy2}
\eqn
s.t.,
\bqs
q(0,t)&\leq& d(0^-,t),\\
q(0,t)&\leq& s(0^+,t),\\
q(0,t)&\leq& C_*, \m{ when } d(0^-,t)>s(0^+,t).
\eqs

Thus we obtain a new LWR model with capacity-drop effect: \refe{inhlwr} with \refe{ctm-entropy} at $x\neq 0$ and \refe{cd-entropy} at $x=0$. The model differs from the traditional LWR model only in the entropy condition at $x=0$. 
We have the following observations regarding the boundary flux function in \refe{cd-entropy}:
\ben
\item When the upstream demand is not greater than the downstream supply, \refe{cd-entropy} is consistent with \refe{ctm-entropy}, and the new LWR model has the same kinematic wave solutions as the traditional one.
\item However, when the upstream demand is greater than the downstream supply and the downstream supply is greater than $C_*$, the capacity drop phenomenon occurs, and the discharging flow-rate is bounded by $C_*$. 
\item The flux function \refe{cd-entropy} is discontinuous in both upstream demands and downstream supplies. This is different from many existing flux functions used in CTM \citep{tampere2011generic,jin2012network}.
\item The new LWR model with \refe{cd-entropy} is purely phenomenological with an exogenous parameter $C_*$, and the driving behaviors and related mechanisms for capacity drop cannot be explained by the model. The model can only be used to describe kinematic waves caused by capacity drop at the lane-drop bottleneck.
\een

\subsection{The Riemann problem}
In this subsection, we show that the new LWR model \refe{inhlwr} is well-defined with the new entropy condition \refe{cd-entropy} at $x=0$ by demonstrating that the Riemann problem has a unique solution with the following initial condition:
\bqs
k(x,0)&=&\cas{{ll}k_1, &x<0;\\k_2, &x>0.} 
\eqs
As for other systems of hyperbolic conservation laws, solutions to the Riemann problem for \refe{inhlwr} at the capacity-drop bottleneck are of physical, analytical, and numerical importance: physically, they can be used to analyze traffic dynamics caused by capacity drop; analytically, \refe{inhlwr} is well-defined if and only if the Riemann problem is uniquely solved \cite{bressan2000convergence}; and numerically, they can be incorporated into the Cell Transmission Model \cite{daganzo1995ctm,lebacque1996godunov}.

Here we solve the Riemann problem by following the analytical framework in \citep{jin2009sd,jin2012_riemann}: (i) the problem is solved in the demand-supply space, with initial conditions:
\bqn
U(x,0)&=&\cas{{ll} (d_1,s_1), &x<0;\\(d_2,s_2), &x>0.} \label{riemannprob}
\eqn
 (ii) in the Riemann solutions on each link, a stationary state arises on a link along with a shock or rarefaction wave, determined by the Riemann problem of the LWR model; (iii) the stationary state should be inside a feasible domain, such that the shock or rarefaction wave propagates backward on the upstream link 1 and forward on the downstream link 2, and the boundary flux $q(0,t)$ equals the stationary flow-rate; (iv) the weak solution space is enlarged to include a filmy interior state on each link at $x=0$, which occupies no space (of measure zero); (v) the entropy condition, \refe{cd-entropy} or \refe{cd-entropy2}, is applied on the interior states; and (vi) we prove that the stationary states and, therefore, the Riemann problem are uniquely solved.

In the demand-supply space, the initial conditions on the upstream and downstream links are denoted by $U_1=(d_1,s_1)$ and $U_2=(d_2,s_2)$, respectively, where $d_i=D(n_i,k_i)$ and $s_i=S(n_i,k_i)$ for $i=1,2$. In the Riemann solutions, upstream stationary and interior states are $U_1^*=(d_1^*,s_1^*)$ and $U_1^0=(d_1^0,s_1^0)$ respectively, and downstream stationary and interior states are $U_2^*=(d_2^*,s_2^*)$ and $U_2^0=(d_2^0,s_2^0)$ respectively. Then the kinematic waves on upstream and downstream links are determined by $RP(U_1,U_1^*)$ and $RP(U_2^*,U_2)$ respectively, which are the Riemann problems for the traditional, homogeneous LWR model. That is, $RP(U_1,U_1^*)$ is the Riemann problem for $\pd{k}t+\pd{Q(n_1,k)}x=0$ with $k(x,0)=\cas{{ll} k_1, &x<0\\ k_1^*, & x>0}$, where $k_1^*=K(n_1,d_1^*/s_1^*)$, and with the traditional Lax entropy condition or the entropy condition in \refe{ctm-entropy}. Since kinematic wave speeds of $RP(U_1,U_1^*)$, $RP(U_1^*,U_1^0)$, $RP(U_2^0,U_2^*)$, $RP(U_2^*,U_2)$ have to be non-positive, positive, negative, and non-negative, respectively, we have the following feasible stationary and interior states \citep{jin2009sd}:
\begin{enumerate}
\item The upstream stationary state is SOC, if and only if $q<d_1$ and $U_1^*=U_1^0=(C_1,q)$; it is UC iff $q=d_1$, $U_1^*=(q,C_1)$, and $s_1^0>d_1$.
\item The downstream stationary state is SUC if and only if $q<s_2$ and $U_2^*=U_2^0=(q,C_2)$; it is OC iff $q=s_2$, $U_2^*=(C_2,q)$, and $d_2^0>s_2$.
\end{enumerate}
We use \refe{cd-entropy2} as an entropy condition in interior states:
\bqn
\max_{U_1^*,U_2^*} q, \label{optimal-entropy}
\eqn
s.t.
\bqs
q&\leq& d_1^0,\\
q&\leq& s_2^0,\\
q&\leq& C_*, \m{ if } d_1^0>s_2^0.
\eqs
The solution of the optimization problem is given by
\bqn
q&=&\cas{{ll} d_1^0, & d_1^0\leq s_2^0\\\min\{s_2^0,C_*\},& d_1^0>s_2^0} \label{lwr:e2}
\eqn
which is consistent with \refe{cd-entropy}.

In the following theorem, we show that the stationary states are uniquely solved with \refe{optimal-entropy}. Furthermore, since one can calculate the boundary flux and the shock or rarefaction waves on both links from the unique stationary states, the Riemann problem is uniquely solved.

\begin{theorem} \label{thm:cd}
For the Riemann problem of \refe{inhlwr} with \refe{ctm-entropy} at $x\neq 0$ and \refe{cd-entropy2} at $x=0$, the stationary states $U_1^*$ and $U_2^*$ and, therefore, the kinematic waves on links 1 and 2 exist and are unique. That is, the optimization problem \refe{optimal-entropy} has a unique solution in $q$, $U_1^*$, and $U_2^*$. In particular,
\bqn
q&=&\cas{{ll}d_1, & d_1\leq s_2\\ \min\{s_2,C_*\}, &d_1>s_2},
\eqn
which is the same as \refe{cd-entropy}. Therefore, the new flux function \refe{cd-entropy} is invariant in the sense of \citep{lebacque2005network,jin2012_riemann}.
\end{theorem}

The proof of Theorem \ref{thm:cd} is given in Appendix A. 
Similar to the inhomogeneous LWR model without capacity drop, the capacity-drop model can have two waves on the two links simultaneously; in contrast, the homogeneous LWR model can only have one wave solution for the Riemann problem. However, the capacity-drop model with  \refe{cd-entropy2} at $x=0$ is different from the non-capacity-drop model with \refe{ctm-entropy} at $x=0$ when $C_*<s_2\leq C_2$ and $s_2<d_1\leq C_1$: in the capacity drop model, $q=C_*$, $U_1^*=(C_1,C_*)$, $U_2^*=(C_*,C_2)$, a backward shock or rarefaction wave forms on the upstream link, and a forward shock or rarefaction wave forms on the downstream link; but in the non-capacity drop model, $q=s_2$, $U_1^*=(C_1,s_2)$, $U_2^*=(C_2,s_2)$, a backward shock or rarefaction wave forms on the upstream link, and a forward rarefaction or no wave forms on the downstream link. That is, when capacity drop occurs, the flow-rate is dropped, and the downstream traffic becomes strictly under-critical. 

\begin{figure} \bc
\includegraphics[width=6in]{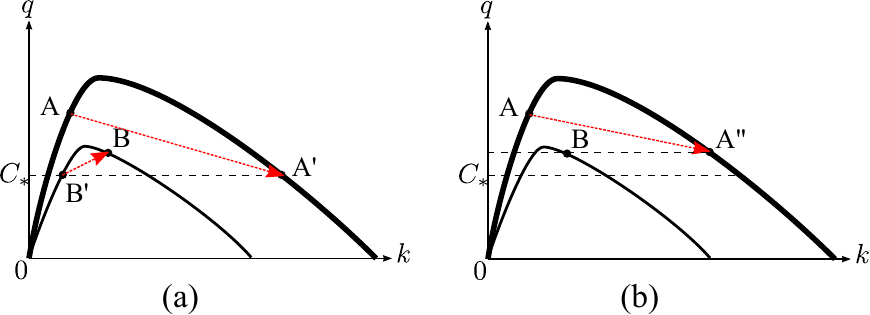}\caption{Kinematic wave solutions of \refe{inhlwr} with an initial upstream condition at $A$ and an initial downstream condition at $B$: (a) In the capacity-drop model with \refe{cd-entropy2} at $x=0$; (b) In the traditional inhomogeneous LWR model with \refe{ctm-entropy} at $x=0$}\label{capacity_drop_not}
\ec
\end{figure}

Consider the example shown in \reff{capacity_drop_not}, where the initial upstream and downstream states are at $A$ and $B$, respectively,  $C_*<s_2<C_2$, and $s_2<d_1<C_1$. In solutions to the capacity-drop model shown in \reff{capacity_drop_not}(a), the stationary states on the upstream and downstream links become $A'$ and $B'$, respectively; the boundary flux becomes $C_*$, which is smaller than the flow-rate of $B$; a backward shock wave forms on the upstream link, and a forward shock wave forms on the downstream link. In solutions to the model without capacity drop shown in \reff{capacity_drop_not}(b), the stationary state on the upstream link becomes $A''$, but the stationary state on the downstream link is the same as the initial state $B$; the boundary flux equals the flow-rate of $B$; a backward shock wave forms on the upstream link, but there is no wave on the downstream link.

\section{Analytical properties of the LWR model of capacity drop}
In this section, we further discuss analytical properties of the LWR model \refe{inhlwr} with the discontinuous entropy condition \refe{cd-entropy} at the lane-drop bottleneck.

\subsection{Stability subject to perturbations in initial and boundary conditions}
We first study the stability of the capacity-drop model, \refe{inhlwr} with \refe{cd-entropy} at $x=0$, subject to perturbations to initial conditions. In particular, we consider solutions of the following perturbed Riemann problem \citep{liu1987relaxation,mascia1997perturbed}:
\bqn
k(x,0)&=&\cas{{ll} k_1, &x<-L\\k_0, &-L<x<0\\ k_2, & x>0} \label{perturbedriemann}
\eqn
where a perturbation $k_0$ is applied on the upstream road section between $-L$ and 0. We expect that results will be similar if we apply a perturbation on the downstream link.
Note that the LWR model \refe{inhlwr} with entropy condition \refe{ctm-entropy} is always stable with respect to perturbations to initial conditions. 

We denote the demand and supply corresponding to $k_i$ by $(d_i,s_i)$ ($i=0,1,2$). 
One can show that, when $d_1<\min\{C_*,s_2\}$ or $d_1>s_2$, solutions with initial condition \refe{perturbedriemann} are the same as those with initial condition \refe{riemannprob} at a large time $t>0$. That is, under these initial conditions, the LWR model \refe{inhlwr} with entropy condition \refe{cd-entropy} is stable subject to perturbations $k_0$. 

\begin{figure} \bc
\includegraphics[width=6in]{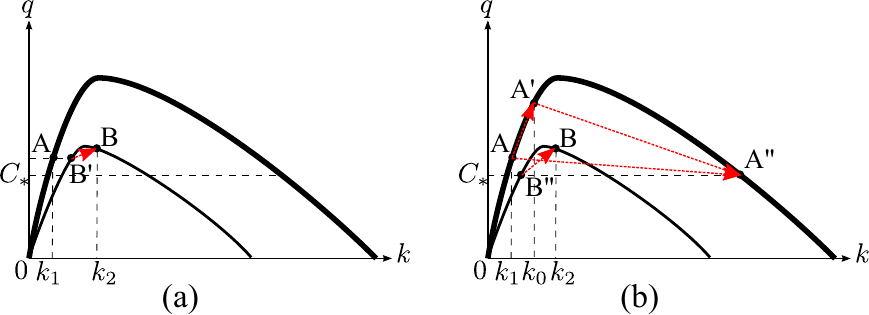}\caption{Kinematic wave solutions of \refe{inhlwr} with an initial upstream condition at $A$ and an initial downstream condition at $B$ in the capacity-drop model: (a) without perturbations, (b) with perturbations}\label{perturbed_rp}
\ec
\end{figure}

However, as shown in \reff{perturbed_rp}, when $C_*<d_1\leq s_2$, solutions to the perturbed Riemann problem can be different from those to the un-perturbed Riemann problem.
In the un-perturbed Riemann problem, both links carry free flow with a flow-rate $q=d_1$, and capacity drop does not occur. However, if a small perturbation leads to an intermediate $d_0>s_2$, capacity drop occurs, a backward shock or rarefaction wave connecting $U_0$ to $(C_1,C_*)$ initiates at $x=0$, and a forward or backward shock wave connecting $U_1$ to $U_0$ initiates at $x=-L$. When the downstream wave connecting $U_0$ to $(C_1,C_*)$ catches up the upstream one connecting $U_1$ to $U_0$, a new shock wave connecting $U_1$ to $(C_1,C_*)$ forms and propagates upstream. In this case, a sufficiently large perturbation to the initial condition can result in totally different solutions. However, if the perturbation is too small such that $d_0\leq s_2$, capacity drop still does not occur. Therefore, the LWR model \refe{inhlwr} with entropy condition \refe{cd-entropy} is bistable in this case. 

\begin{figure} \bc
\includegraphics[width=6in]{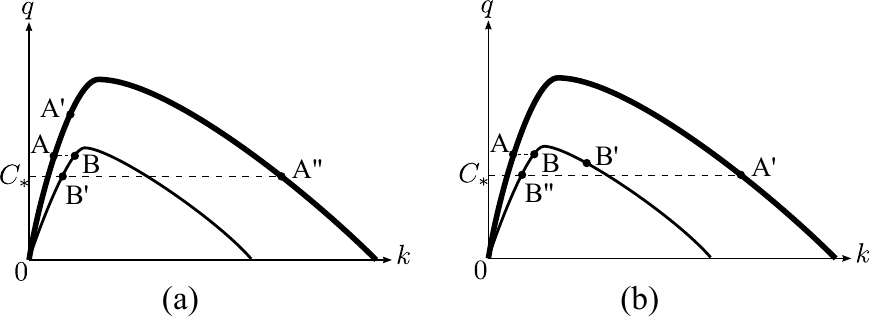}\caption{Activation of capacity drop: (a) A high-density platoon on the upstream link; (b) A congested queue on the downstream link}\label{breakdown}
\ec
\end{figure}

When the road with a lane drop in \reff{lane-drop} carries free flow with a flow-rate greater than $C_*$, traffic breakdown and capacity drop can also be induced by oscillations in both upstream demand and downstream supply. We demonstrate the process in \reff{breakdown}. If initially the upstream link carries a uniform, free flow traffic at $(d_1,C_1)$ (point $A$ in \reff{breakdown}) and the downstream link carries a uniform, free flow traffic at $(d_1,C_2)$ (point $B$ in \reff{breakdown}), where $C_*<d_1\leq C_2$. 
\ben
\item If a platoon of vehicles from the upstream link, which has a high density with a demand greater than $C_2$ (point $A'$ on \reff{breakdown}(a)), reaches the lane-drop bottleneck, then vehicles queue up on the upstream link, capacity drop is activated, and traffic on the upstream link breaks down and becomes $(C_1,C_*)$ (point $A''$ in \reff{breakdown}(a)). Correspondingly, traffic on the downstream link becomes $(C_*,C_2)$ (point $B'$ in \reff{breakdown}(a)). The throughput drops from $d_1$ to $C_*$.
\item If a congested queue, which has a supply smaller than $d_1$ (point $B'$ in \reff{breakdown}(b)), propagates to the lane-drop area, then vehicles queue up on the upstream link, capacity drop is activated, and traffic on the upstream link breaks down and becomes $(C_1,C_1^*)$ (point $A'$ in \reff{breakdown}(b)). Correspondingly, traffic on the downstream link becomes $(C_*,C_2)$ (point $B"$ in \reff{breakdown}(b)). The throughput drops from $d_1$ to $C_*$. 
\een
Both scenarios confirm that the new kinematic wave model replicates the two main characteristics of capacity drop: (i) capacity drop occurs with an upstream queue, and (ii) the throughput drops once it is activated.
In addition, as observed in real world \citep{persaud1998exploration,persaud2001breakdown}, the capacity drop as well as traffic breakdown can be induced by random fluctuations in upstream and downstream conditions even when the upstream is uncongested but carries a flow-rate higher than the dropped capacity, $C_*$. 

\subsection{Discontinuous flow-density relation in stationary states}

In this subsection, we consider the following traffic statics problem on a road section $x\in[-X,Y]$ with a lane-drop at $x=0$. Initially the road section is empty with $k(x,0)=0$. The upstream demand is constant, $d(-X^-,t)=d_0$, and the downstream supply is also constant, $s(Y^+,t)=s_0$. We are interested in finding stationary states in the road network \citep{jin2012statics}.

In stationary states, both the upstream and downstream links carry uniform traffic\footnote{Here we do not consider zero-speed shock waves on a link as in \citep{jin2012statics}.}, and we assume that their densities are $k_1$ and $k_2$, respectively. Then the corresponding demands and supplies are $(d_1,s_1)$ and $(d_2,s_2)$, respectively. We denote the flow-rate in the network by $q$, which is constant at all locations.
Then using \refe{cd-entropy} at the lane-drop location and \refe{ctm-entropy} at the origin and destination, we have
\bsq
\bqn
q&=&\min\{d_0,s_1\},\\
q&=&\cas{{ll} d_1, &d_1\leq s_2\\ \min\{s_2,C_*\},&d_1>s_2}\\
q&=&\min\{d_2,s_0\}.
\eqn
In addition, from the definitions of supply and demand we have
\bqn
C_1&=&\max\{d_1,s_1\},\\
C_2&=&\max\{d_2,s_2\}.
\eqn
\esq

From the five equations above and the evolution of traffic dynamics \footnote{The evolution of traffic dynamics can be analyzed with shock and rarefaction waves, but the detailed analysis is omitted.}, we can find the following solutions of $(d_1,s_1)$ and $(d_2,s_2)$: \footnote{Without loss of generality, we assume that $d_0\leq C_1$ and $s_0\leq C_2$.}
\ben
\item When $d_0\leq  s_0\leq C_2$, $q=d_0$, $(d_1,s_1)=(d_0,C_1)$, and $(d_2,s_2)=(d_0,C_2)$. In this case, both links carry free flow.
\item When $d_0>s_0$ and $s_0\leq C_*$, $q=s_0$, $(d_1,s_1)=(C_1,q)$, and $(d_2,s_2)=(C_2,q)$. In this case, both links carry congested traffic.
\item When $d_0>s_0$ and $s_0>C_*$, $q=C_*$, $(d_1,s_1)=(C_1,q)$, and $(d_2,s_2)=(q,C_2)$. In this case, link 1 is congested, but link 2 is not.
\een
Note that, if the initial densities are not zero, other types of stationary states can exist.

\begin{figure}\bc
\bc $\ba{c@{\hspace{0.3in}}c}
\includegraphics[height=1.8in]{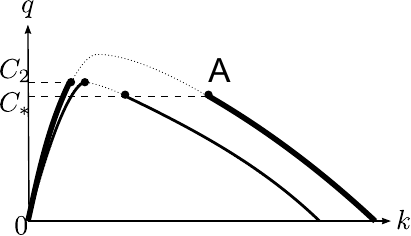} &
\includegraphics[height=1.8in]{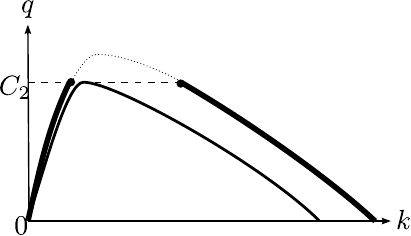} \\
\multicolumn{1}{c}{\mbox{\bf (a) } \m{With capacity drop}} &
    \multicolumn{1}{c}{\mbox{\bf (b) } \m{Without capacity drop}}
\ea$ \ec \caption{Stationary flow-density relations of upstream and downstream links in a lane drop area: thinner solid curves for the downstream link, and thicker solid curves for the upstream link}\label{fd_cd_ss} 
\ec
\end{figure}

Then from the relationship between congestion level and density, we can find corresponding densities and therefore fundamental diagrams in stationary states on both upstream and downstream links, shown in \reff{fd_cd_ss}(a). From the figure, we can see that the stationary flow-density relations are discontinuous in both the upstream and downstream parts, even though the original fundamental diagrams are continuous. This discontinuity is caused by the new entropy condition of capacity drop, \refe{cd-entropy}. In comparison, the stationary flow-density relations without capacity drop are shown in \reff{fd_cd_ss}(b), from which we can see that the downstream flow-density relation is continuous, but the upstream one is still discontinuous, since higher flow-rates cannot be sustained due to the lane-drop bottleneck.

From \reff{fd_cd_ss} we can see that, even though the fundamental diagrams are still continuous, only discontinuous portions of them are observable due to lane drop and capacity drop. 
Therefore, such ``discontinuous'' fundamental diagrams should be called incomplete fundamental diagrams.
Such incomplete fundamental diagrams are consistent with many observations, e.g., Figures 12 and 15 of \citep{drake1967fd}, the schematic Figure 2 of \citep{hall1992fd}, and Figure 6 for shoulder lane in \citep{hall1986fd}. This can also be used to explain Figures 2 and 3 in \citep{hall1991capacity}, since the downstream is never congested. 
But they are not consistent with \citep{koshi1983fd}, which is very scattered and may include non-stationary states \citep{cassidy1998bivariate}.
This is also consistent with the conjecture that discontinuous fundamental diagrams could be caused by capacity drop at active bottlenecks \citep{hall1991capacity}.
However, incomplete fundamental diagrams are the effect of capacity drop, but capacity drop is not the effect of incomplete fundamental diagrams. In addition, capacity drop is not the only cause of incomplete flow-density relations.

\section{Capacity drop in a ring road}
\begin{figure}
\bc
\includegraphics[height=2in]{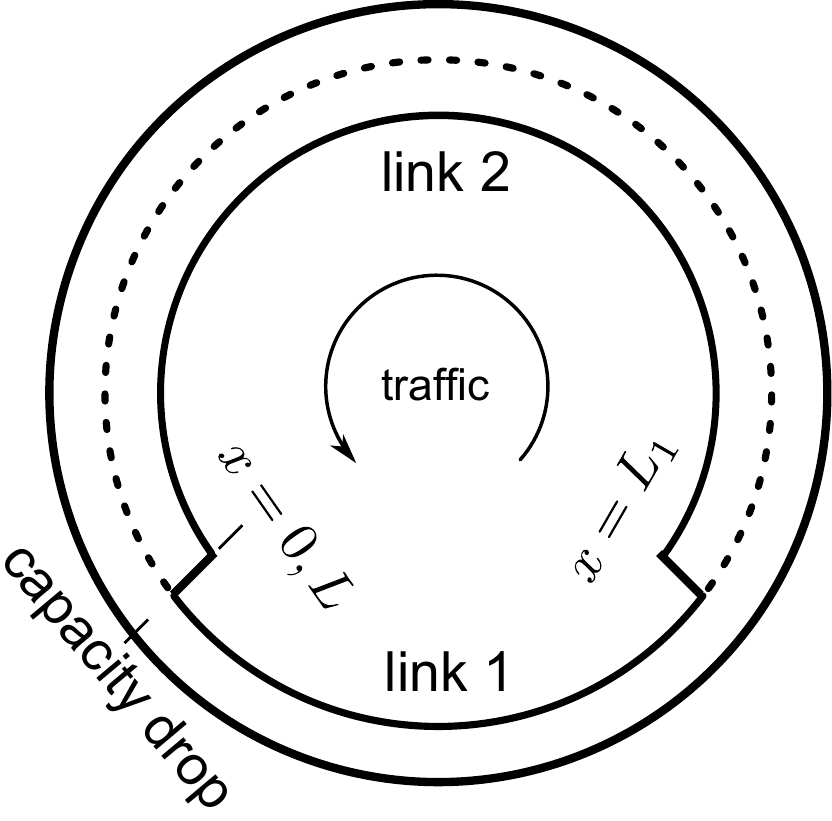}
\ec
\caption{An inhomogeneous ring road with capacity drop}\label{inhomogeneous_ring_cd}
\end{figure}

In this section we solve the LWR model \refe{inhlwr} on the inhomogeneous ring road with a length of $L$ shown in \reff{inhomogeneous_ring_cd}, in which the traffic direction is shown by the arrow. The ring road is composed of two homogeneous links, whose capacities are $C_1$ and $C_2>C_1$, respectively. We assume that the fundamental relationships for two links are given by $q=Q_1(k)$ and $k=K_1(\gamma)$ for $x\in[0,L_1]$, and $q=Q_2(k)$ and $\r=K_2(\gamma)$ for $x\in[L_1,L]$. 
In addition, we assume that capacity drop occurs at $x=0$ or $x=L$, and the dropped capacity is $C_*<C_2$. 
That is, \refe{ctm-entropy} is used at any location except at $x=0$ or $x=L$, where \refe{cd-entropy} is used.

The discrete CTM with capacity drop can be developed for the LWR model \refe{inhlwr} as follows. We first divide both links into cells and discretize a simulation duration into time steps. The cell length $\dx$ and the time-step size $\dt$ should satisfy the CFL condition \citep{courant1928CFL}, such that a vehicle cannot traverse a whole cell during a time interval.
At time step $j$, traffic density in cell $i$ ($i=1,\cdots,N$) is denoted by $k_i^j$, and the corresponding demand and supply by $d_i^j$ and $s_i^j$, respectively. Then from traffic conservation we have the following equation to update traffic density in a cell:
\bqs
k_i^{j+1}&=&k_i^j+\frac{\dt}{\dx}(q_{i-1/2}^{j}-q_{i+1/2}^j),
\eqs
where $q_{i-1/2}^j$ is the boundary flux from cell $i-1$ to cell $i$ during $[j\dt,(j+1)\dt]$. If capacity drop occurs at the boundary between cell $i-1$ and cell $i$, we use \refe{cd-entropy} to calculate the boundary flux
\bqs
q_{i-1/2}^{j}&=&\cas{{ll} d_{i-1}^j, &d_{i-1}^j\leq s_i^j\\\min\{s_i^j,C_*\},& d_{i-1}^j> s_i^j}.
\eqs
Otherwise, we use \refe{ctm-entropy} to calculate the boundary flux $q_{i-1/2}^j =\min\{d_{i-1}^j, s_i^j\}$.
When $j=0$, initial densities $k_i^0$ and, therefore, initial demands and supplies are given.

\subsection{Stationary states and macroscopic fundamental diagram}
In this subsection we consider possible stationary states on the ring road. That is, traffic density is time-independent at any location. On link $i$, $\pd{k_i} t=0$ for $x\in[0,L]$, and 
\bqs
\der{q}t&=&\pd{q}x \der{x}t+\pd{q} t=(Q_i'(k_i)-\der xt)\pd {k_i}t=0.  
\eqs
Therefore $q(x,t)=q$ is constant on the ring road. Inside a homogeneous link $i$, traffic can be stationary at UC ($k_i(x,t)=K_i(q/C_i)$), SOC ($k_i(x,t)=K_i(C_i/q)$), or a zero-speed shock wave (ZS) connecting an upstream SUC state ($k_i(x,t)=K_i(q/C_i)$) and a downstream SOC state ($k_i(x,t)=K_i(C_i/q)$). Then all possible combinations of stationary states are explained in the following.
\bi
\item When link 1 is stationary at UC with $q \leq C_1$,  $d_1(L_1^-,t)=q$ and $s_1(0^+,t)=C_1$. Then we have the following scenarios. (a) Link 2 can be stationary at UC with   $d_2(x,t)=q$ and $s_2(x,t)=C_2$ for $x\in(L_1,L)$, and the total number of vehicles on the ring road is $N_a=K_1(q/C_1) L_1+K_2(q/C_2) (L-L_1)$. (b) Link 2 can be stationary at ZS with $d_2(L^-,t)=C_2$ and $s_2(L_1^+,t)=C_2$, and we have that $q=C_*$. Assuming that link 2 is SUC for $x\in[L_1,L_2]$ and SOC for $x\in[L_2,L]$. In this case, the total number of vehicles on the ring road is $N_b=K_1(C_*/C_1) L_1+K_2(C_*/C_2) (L_2-L_1)+K_2(C_2/C_*) (L-L_2)$, which varies with respect to $L_2$.  (c) Link 2 can also be stationary at SOC with $d_2(L^-,t)=C_2$ and $s_2(L_1^+,t)=q$, and we have $q=C_*$. In this case, the total number of vehicles on the ring road is $N_c=K_1(C_*/C_1) L_1+K_2(C_2/C_*) (L-L_1)$.
\item When link 1 is stationary at ZS with $q<C_1$, $d_1(L_1^-,t)=C_1$, and $s_1(0^+,t)=C_1$. Then we have the following scenario. (d) Link 2 can be stationary at SOC with $d_2(L^-,t)=C_2$ and $s_2(L_1^+,t)=q=C_*$. Assuming that link 1 is SUC for $x\in[0,L_0]$ and SOC for $x\in[L_0,L_1]$. In this case, the total number of vehicles on the ring road is $N_d= K_1(C_*/C_1) L_0+ K_1(C_1/C_*) (L_1-L_0)+ K_2(C_2/C_*) (L-L_1)$.
It can be verified that link 2 cannot be stationary at UC or ZS, since, otherwise, $s_2(L_1^+,t)=C_2$, and we have $q=\min\{C_2,C_1\}=C_1$, which contradicts $q<C_1$.
\item When link 1 is stationary at SOC with $q<C_1$, $d_1(L_1^-,t)=C_1$, and $s_1(0^+,t)=q$. Then we have the following scenario. (e) Link 2 can be stationary at SOC with $d_2(L^-,t)=C_2$ and $s_2(L_1^+,t)=q$, if $q\leq C_*$. In this case, the total number of vehicles on the ring road is $N_e=K_1(C_1/q)L_1+K_2(C_2/q) (L-L_1)$. It can be verified that link 2 cannot be stationary at UC or ZS, since, otherwise, $s_2(L_1^+,t)=C_2$, and we have $q=\min\{C_2,C_1\}=C_1$, which contradicts $q<C_1$.
\ei

If we define traffic density of the whole network by $k=N/L$, then we can obtain a macroscopic fundamental diagram $q=Q(k)$ for the five scenarios:
\bi
\item[(a)] $q\leq C_1$ and $k=K_1(\frac{q}{C_1})\frac{L_1}L+K_2(\frac{q}{C_2})(1-\frac{L_1}L)$.
\item[(b)] $q=C_*$ and $k=K_1(\frac{C_*}{C_1}) \frac{L_1}L+K_2(\frac{C_*}{C_2}) \frac{L_2-L_1}L+K_2(\frac{C_2}{C_*}) (1-\frac{L_2}L)$, where $L_1<L_2<L$.
\item[(c)] $q=C_*$ and $k=K_1(\frac{C_*}{C_1}) \frac{L_1}L+K_2(\frac{C_2}{C_*}) (1-\frac{L_1}L)$.
\item[(d)] $q=C_*$ and $k=K_1(\frac{C_*}{C_1}) \frac{L_0}L+K_1(\frac{C_1}{C_*}) \frac{L_1-L_0}L+K_2(\frac{C_2}{C_*}) (1-\frac{L_1}L)$, where $0<L_0<L_1$.
\item[(e)] $q\leq C_*$ and $k=K_1(\frac{C_1}{q})\frac{L_1}L+K_2(\frac{C_2}{q})(1-\frac{L_1}L)$.
\ei

\begin{figure}
\bc
\includegraphics[width=5in]{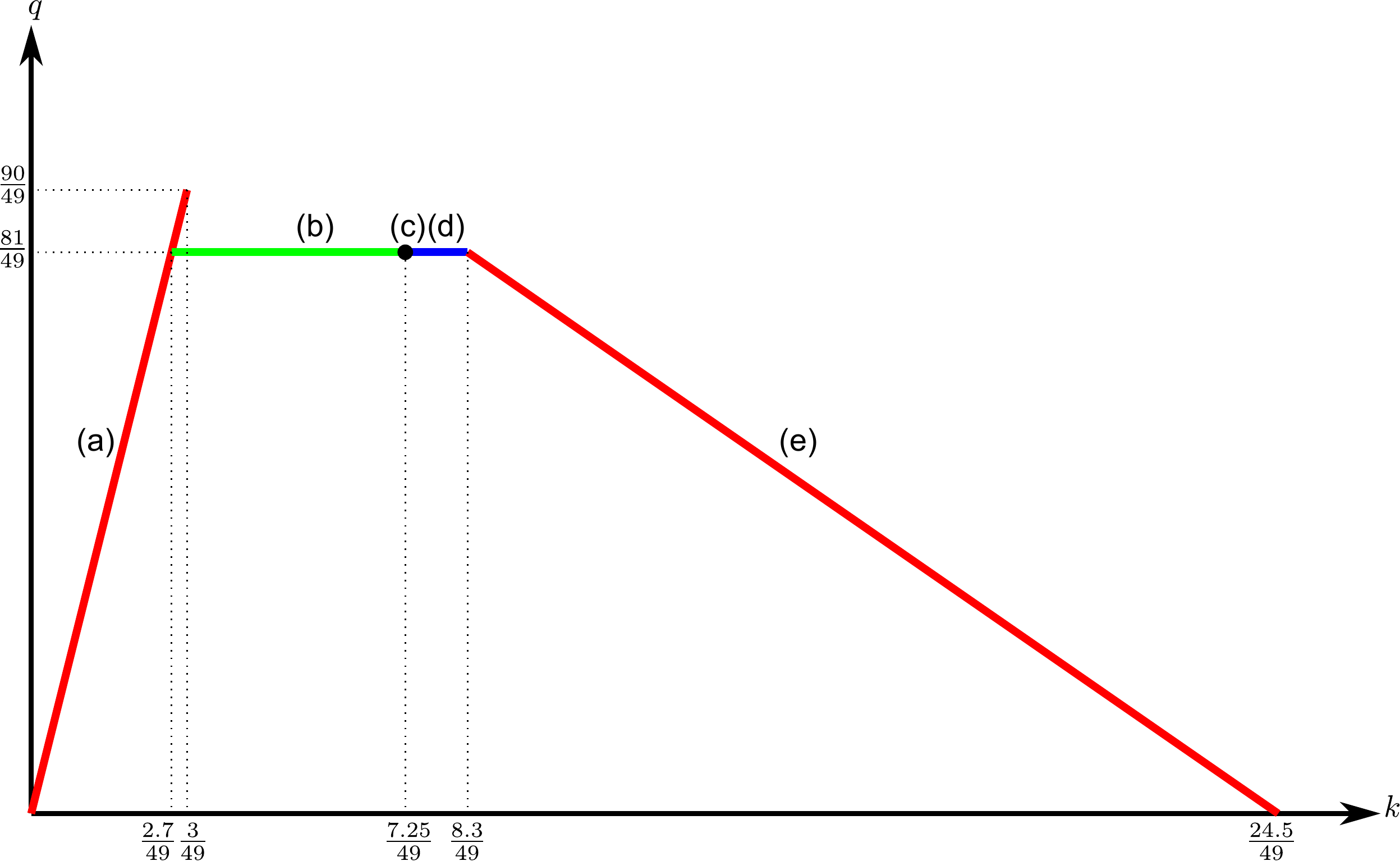}\caption{A macroscopic fundamental diagram in an inhomogeneous ring road with capacity drop} \label{inhring_fd}
\ec
\end{figure}

As an example, we use the triangular fundamental diagram \refe{tri-fd} for the two links with $n_1=3$ and $n_2=4$. In addition, we set $v^*=30$ m/s, $k^*=\frac 17$ veh/m, and $\tau=1.4$ s. Then $k_c(n)=\frac{n}{49}$, $q_C(n)=\frac{30 n}{49}$, and 
\bqs
\gamma&=&\cas{{ll}\frac{49k}n, & k\leq \frac{n}{49}\\\frac{6n}{7n-49k},&k>\frac{n}{49}}\\
k&=&K(n,\gamma)=\cas{{ll} \frac{n\gamma}{49},& \gamma\leq 1\\ \frac n7-\frac{6n}{49\gamma}, & \gamma>1}
\eqs
We assume that the dropped capacity is $C_*=0.9 q_C(3)=\frac{81}{49}$ veh/s.
Thus, we have
\bi
\item[(a)] $q\leq \frac{90}{49}$ and $k=\frac{q}{30}$.
\item[(b)] $q=\frac{81}{49}$ and $k=\frac{2.7}{49}\frac{L_2}L+\frac{11.8}{49}(1-\frac{L_2}L)$, where $L_1<L_2<L$.
\item[(c)] $q=\frac{81}{49}$ and $k=\frac{2.7}{49}\frac{L_1}L+\frac{11.8}{49}(1-\frac{L_1}L)$.
\item[(d)] $q=\frac{81}{49}$ and $k=\frac{2.7}{49}\frac{L_0}L+\frac{4.8}{49}\frac{L_1-L_0}L+\frac{11.8}{49}(1-\frac{L_1}L)$, where $0<L_0<L_1$.
\item[(e)] $q\leq\frac{81}{49}$ and $k=\frac 47-\frac 17 \frac{L_1}L-\frac q5$.
\ei
Therefore, the macroscopic fundamental diagram is given by
\bqs
q&=&\cas{{ll} 30 k, & 0\leq k\leq \frac{3}{49}\\ \frac{81}{49}, & \frac{2.7}{49}<k<\frac{11.8}{49}-\frac 17\frac{L_1}L \\ \frac{20}7-\frac 57 \frac{L_1}L-5k, & \frac{11.8}{49}-\frac 17\frac{L_1}L\leq k\leq \frac47 -\frac 17 \frac{L_1}L}
\eqs
In particular, if $L_1=\frac 12 L$; i.e., if links 1 and 2 have the same length, the macroscopic fundamental diagram is shown in \reff{inhring_fd}. \footnote{We can show that, without capacity drop, the macroscopic fundamental diagram for the ring road is of a trapezoidal shape.}
From the figure, we can see a discontinuity in the macroscopic fundamental diagram consistent with the schematic Figure 2 of \citep{hall1992fd}. 
In addition, when $\frac{2.7}{49}\leq k\leq \frac{3}{49}$, $q$ can have two values: one for free flows on both links in scenario (a), and the other for UC link 1 and ZS link 2 in scenario (b).
Note that for an open road section with capacity drop, if traffic is stationary on the road section, then the observed fundamental diagram would also look like \reff{inhring_fd}. 

\subsection{Stability and bifurcation of stationary states}

In this subsection, we numerically study traffic dynamics on the inhomogeneous ring road in \reff{inhomogeneous_ring_cd} with $L=1960$ m and $L_1=980$ m. The triangular fundamental diagram \refe{tri-fd} is used for both links with $n_1=3$ and $n_2=4$. As in the preceding subsection, we set $v^*=30$ m/s, $k^*=\frac 17$ veh/m, and $\tau=1.4$ s.
Then capacities for two links are $\frac{90}{49}$ and $\frac{120}{49}$ v/s, respectively. We assume that the  capacity drops by 10\% at the lane-drop location and $C_*=\frac{81}{49}$.

Here we consider the following initial condition:
\bqn
k(x,0)&=&\frac{2.8}{49}+ \epsilon(x,0),\\
\epsilon(x,0)&=&\cas{{ll}\epsilon, &x\in[L-70,L)\\-\epsilon, &x\in [L-140,L-70)\\0,&x\in[0,L-140)}
\eqn
where $\epsilon$ is the oscillation magnitude. That is, we apply a small oscillation on link 2.
Then, the total number of vehicles on the ring road is $N=112$, and average traffic density is $k=\frac{2.8}{49}$. From \reff{inhring_fd} we can see that the ring road can become stationary with both UC links in scenario (a) or with link 1 UC and link 2 ZS in scenario (b).

\begin{figure}
\bc $\ba{c@{\hspace{0.3in}}c}
\includegraphics[height=2.5in]{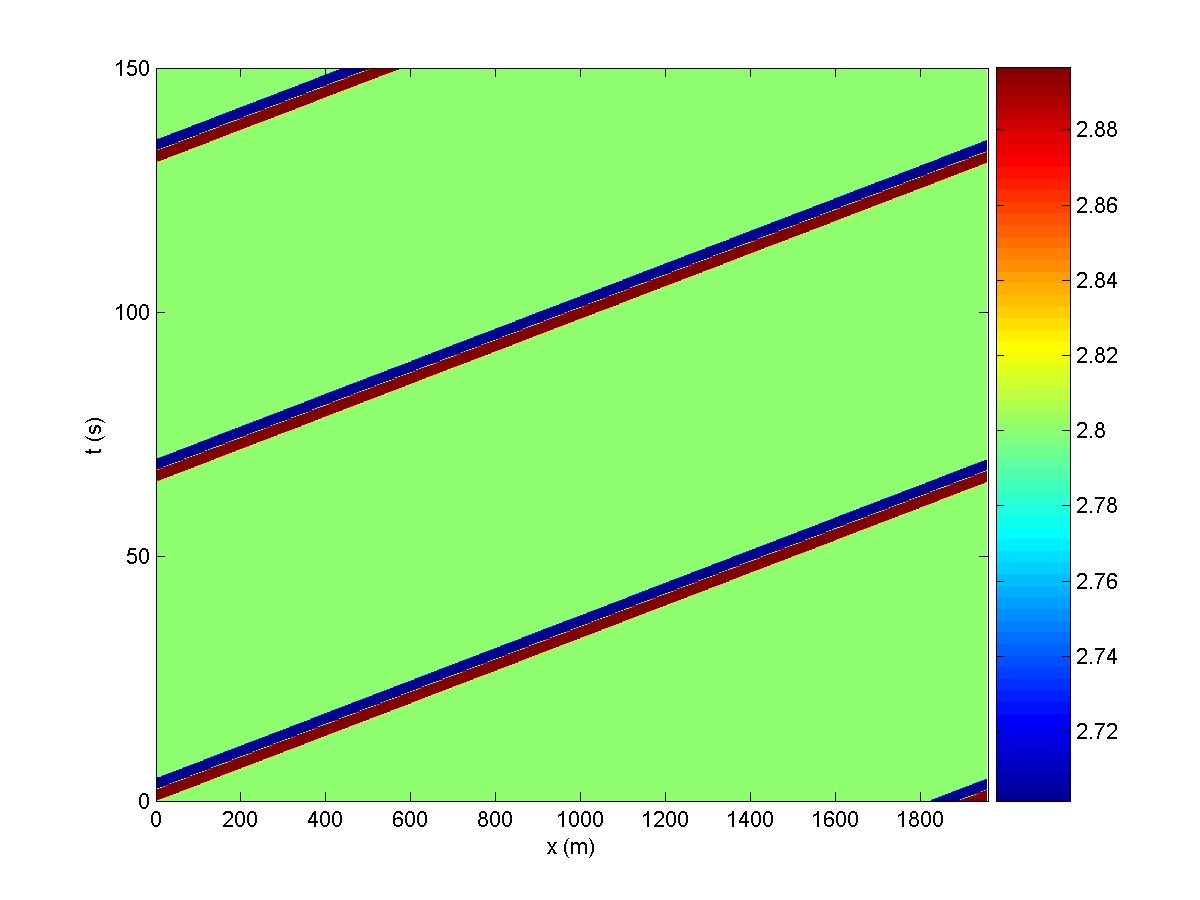} &
\includegraphics[height=2.5in]{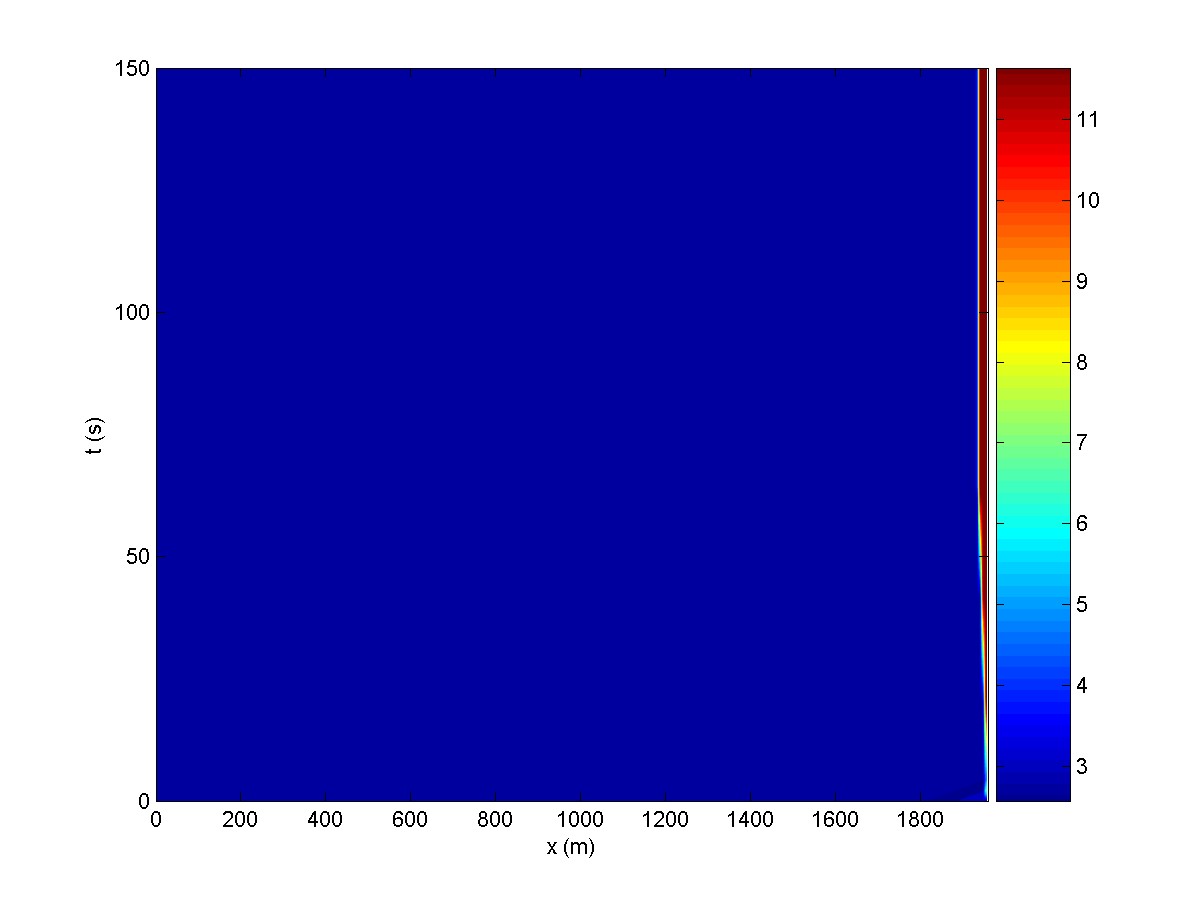} \\
\multicolumn{1}{c}{\mbox{\bf (a) } \epsilon=\frac{0.1}{49} \m{ v/m}} &
    \multicolumn{1}{c}{\mbox{\bf (b) } \epsilon=\frac{0.3}{49} \m{ v/m}}
\ea$ \ec \caption{Contour plots of $49 \cdot k(x,t)$}\label{inh_ring_cd} 
\end{figure}

In the following, we simulate traffic dynamics with the Godunov finite difference equation. We set $\dx=7$ m and $\dt=7/30$ s, which satisfy the CFL condition. The simulation duration is  $T=150$ s. 
When $\epsilon=\frac{0.1}{49}$ and $\frac{0.3}{49}$ v/m, the results are shown in \reff{inh_ring_cd}. From the figure, we can see that, when $\epsilon=\frac{0.1}{49}$ v/m is small, the oscillation does not converge or diverge, and the ring road has an average flow-rate of $\frac{84}{49}$ v/s; but when $\epsilon=\frac{0.3}{49}$ v/m, the ring road converges to a stationary state of type (b) in \reff{inhring_fd}, in which there is a queue on link 2, and the average flow-rate becomes $\frac{81}{49}$ v/s. From more values of $\epsilon$, we can see that, when $\epsilon>\frac{0.2}{49}$ v/m, the ring road converges to a stationary state of type (b). This suggests that the traffic system is bistable with a flow-rate of either $\frac{84}{49}$ or $\frac{81}{49}$ v/s.

\section{An empirical observation of the flow-density relation at an active bottleneck}
In this section we present an empirical observation of the flow-density relation located at the merging section between I-405 South and Jeffrey Road in Irvine, CA. The study site is shown in \reff{fig:I405S-Jeffrey}(a). This location has three vehicle detector stations (VDS's): upstream mainline VDS 1201171, on-ramp VDS 1201165, and downstream mainline VDS 1209189, which are shown as the blue circles in Figure \reff{fig:I405S-Jeffrey}(a). 
We use the detector data from 5:00 AM to 10:00PM on April 4th, 2012 and aggregate the data over lanes. 
In Figures  \reff{fig:I405S-Jeffrey}(c) and (d), we provide 30-second flow-rates and speeds from 4:30 PM to 7:30 PM for both upstream detectors (VDS 1201171 and VDS 1201165) and downstream detectors (VDS 1209189).

\bfg \bc
		$\ba{c@{\hspace{0.1in}}c}
		\includegraphics[width=3in]{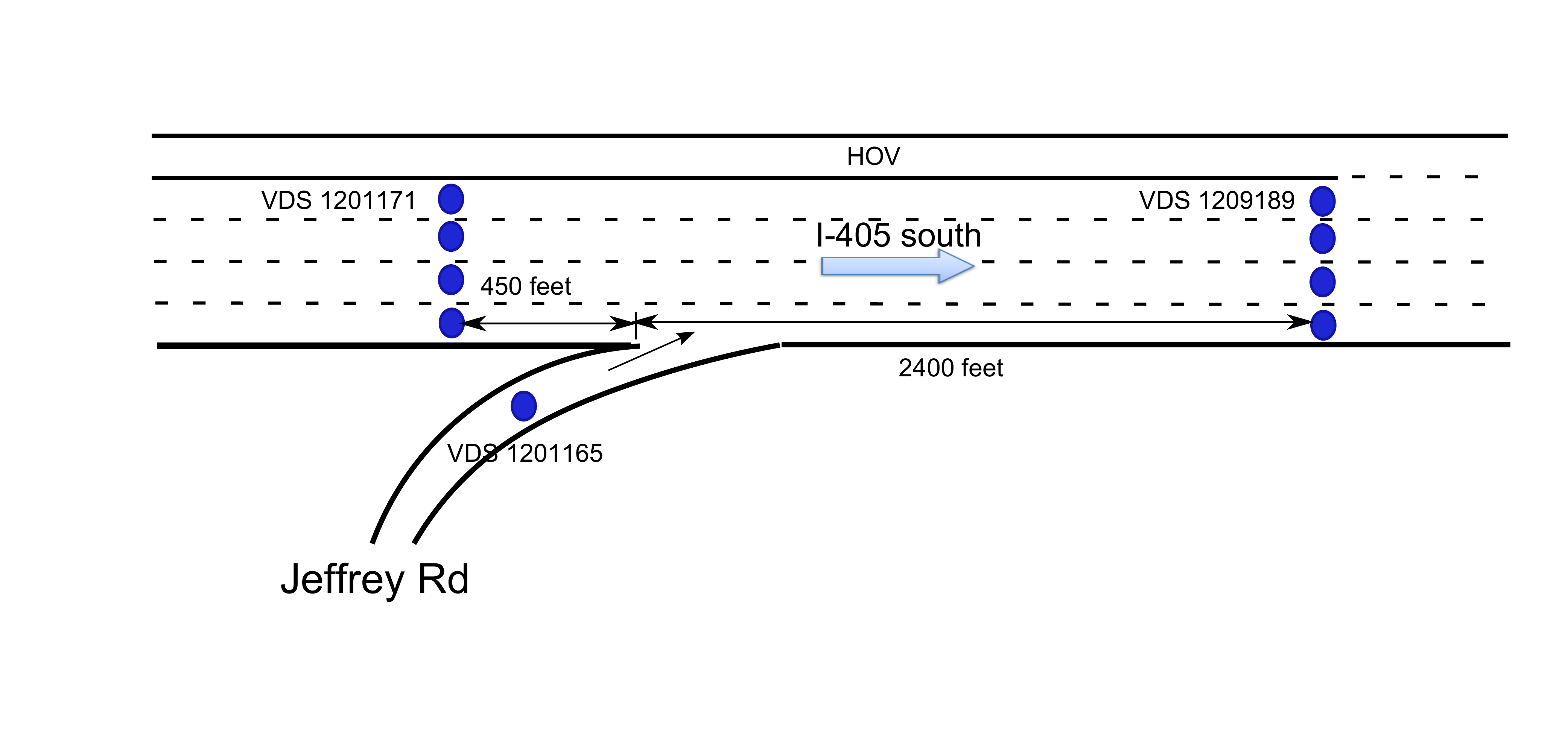} &
		\includegraphics[width=3in]{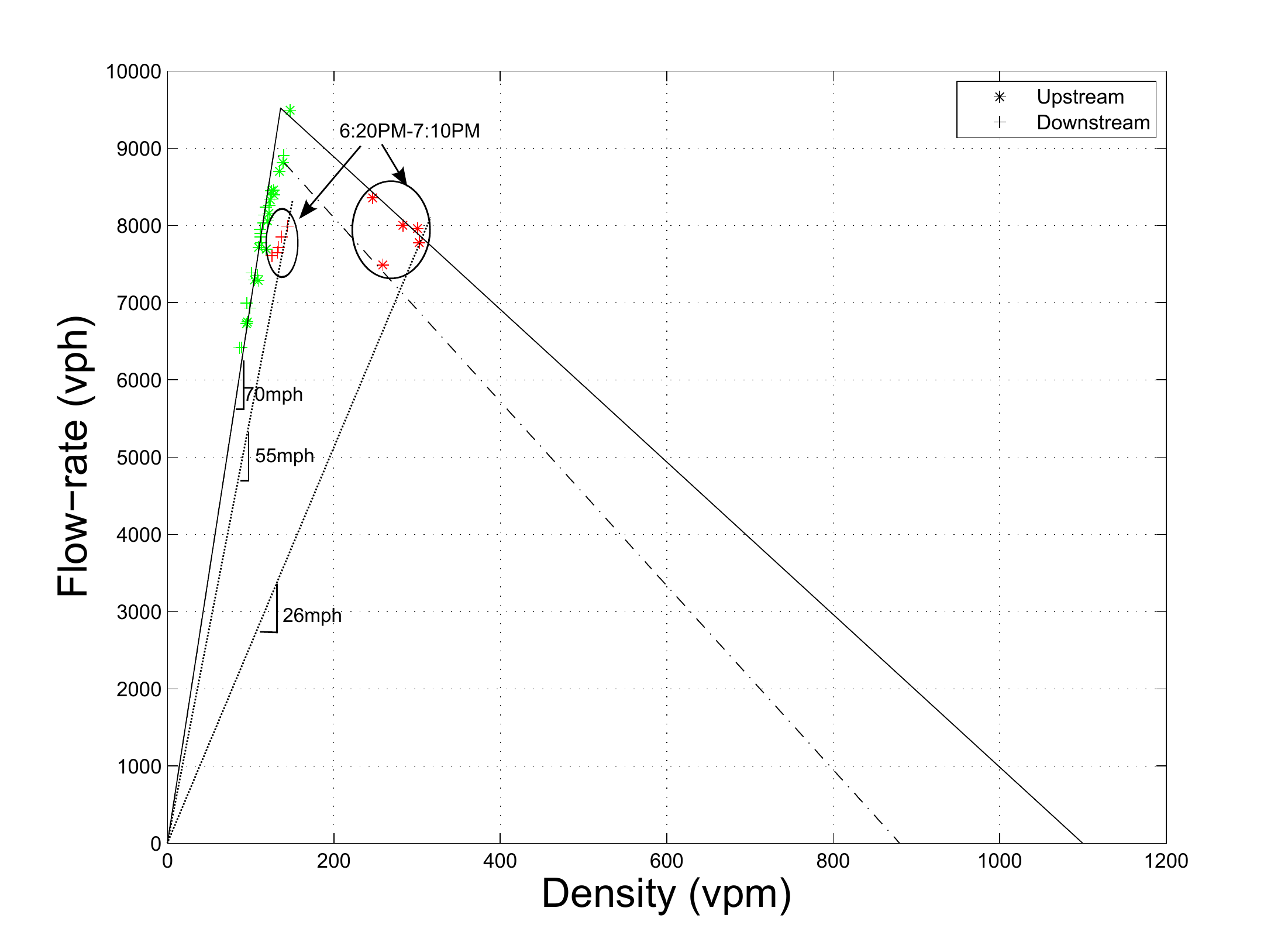} \\
		\multicolumn{1}{c}{\mbox{\bf (a)}} &
		    \multicolumn{1}{c}{\mbox{\bf (b)}}\\
		  \includegraphics[width=3in]{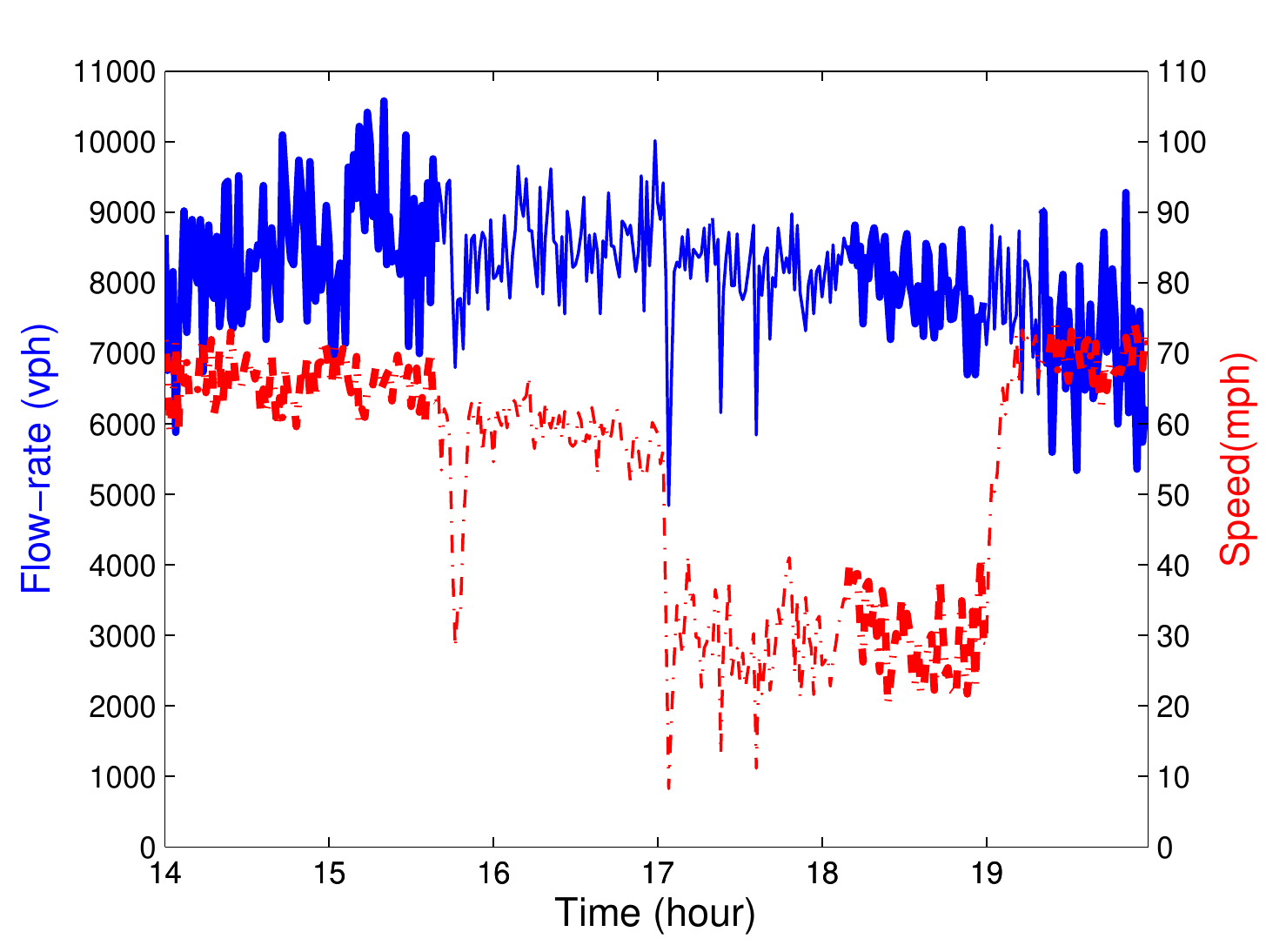} &
		\includegraphics[width=3in]{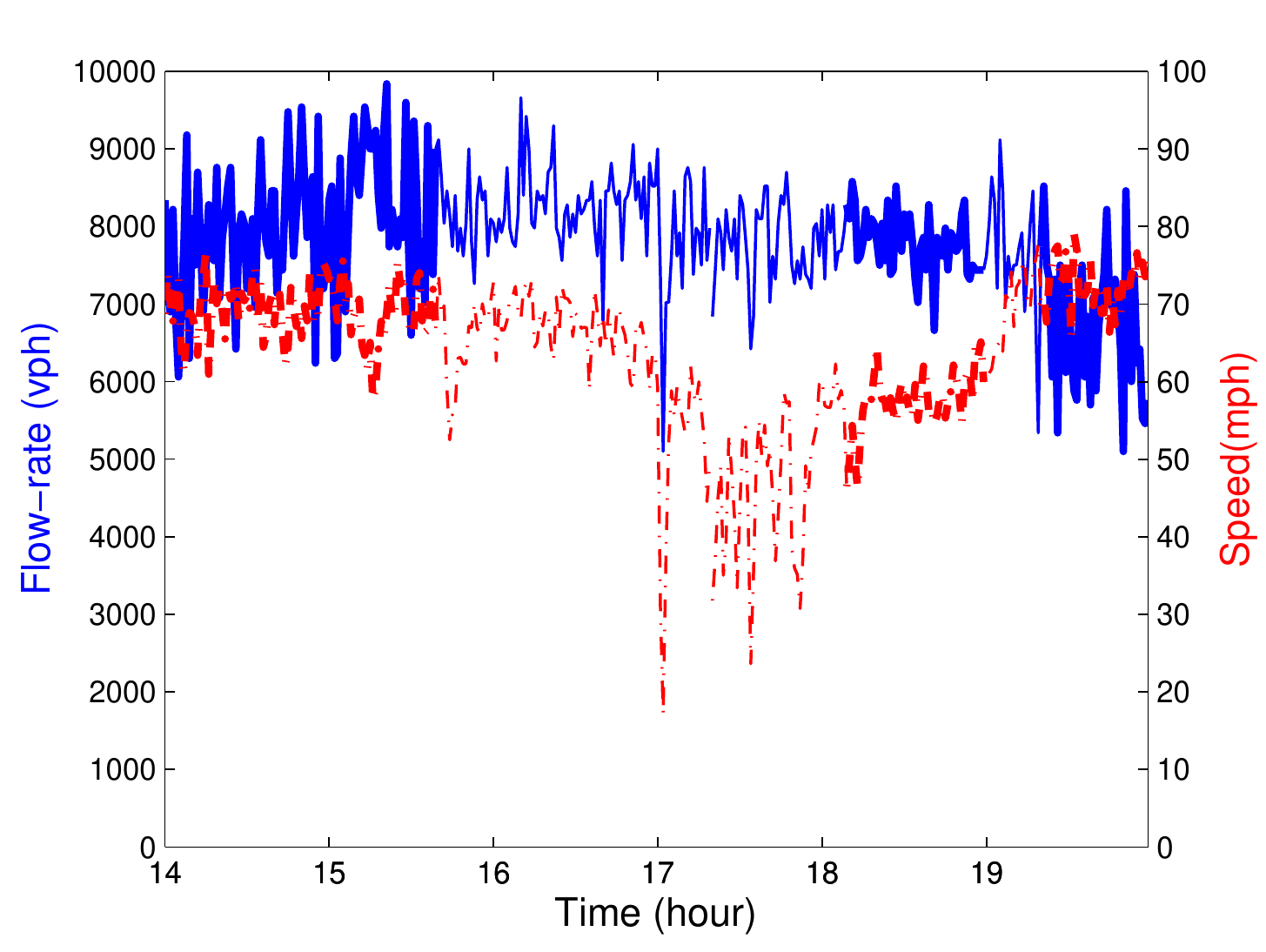} \\
		\multicolumn{1}{c}{\mbox{\bf (c)}} &
		    \multicolumn{1}{c}{\mbox{\bf (d)} }
		\ea$
		\caption{Empirical observation of the flow-density relation: (a)The study site, (b)The flow-density relation, (c) Upstream flow-rates and speeds, and (d) Downstream flow-rates and speeds} \label{fig:I405S-Jeffrey} \ec 
\efg

This location is uncongested except during the afternoon peak hours, when the bottleneck is only activated due to very high on-ramp flow-rates, and the upstream mainline freeway becomes congested. 
Therefore, we expect to have similar observations as in  Figures 2 (upstream) and 3 (downstream) in \citep{hall1991capacity}. Theoretically, the stationary flow-density relations for upstream and downstream links should look alike those in \reff{fd_cd_ss}(a), but we cannot observe the congested parts except state A for congested upstream road. 

In \reff{fig:I405S-Jeffrey}(b), we show the flow-density relations in near-stationary states, which are defined as in \citep{cassidy1998bivariate}. Here we assume there is a linear relation between occupancy and density; i.e., density equal occupancy divided by $100g$, where  the g-factor $g=22$ feet. 
In the figure, the flow-density relations in free-flow are shown by green asterisks (upstream) and plus signs (downstream), and those in congested traffic by red asterisks (upstream) and plus signs (downstream).
Note that, even though the downstream speeds in \reff{fig:I405S-Jeffrey}(d) are also lower than the free-flow speed during 6:20-7:10 PM, there is no congestion at this location. Rather the lower speeds are caused by vehicles accelerating away from the queue between the two detectors. Therefore in \reff{fig:I405S-Jeffrey}(b) we do not observe the congested traffic states at the downstream detector, even though the red plus signs have lower speeds than the green ones. In comparison, we are able to observe congested states at the upstream detector (red asterisks), and these congested states have nearly the same flow-rates around 8000 vph (four lanes). However, in free-flow traffic, the maximum flow-rate can reach 9500 vph at both upstream and downstream detectors. Therefore, capacity drop occurs at this location, and the capacity drop magnitude is about 16\%. This observation verifies the  prediction made based on the new capacity drop model.

\section{Conclusion}
In this paper, we proposed a phenomenological model of capacity drop within the framework of kinematic wave theories. In particular, for capacity drop occurring at a lane drop location, we introduced a new entropy condition, in which the boundary flux is reduced to a dropped capacity when the upstream demand is higher than the downstream supply.
The model is consistent with observations in that capacity drop is activated when vehicles start to queue up at the upstream section. 
We then theoretically showed that the model is well-defined since the Riemann problem is uniquely solved. We also demonstrated that the model leads to discontinuous flow-density relations in stationary traffic, and it is bistable since traffic breakdown and capacity drop can be activated by sufficiently large perturbations in initial and boundary conditions. For a ring road with a lane drop, we analytically derived a macroscopic fundamental diagram consistent with that in literature and with numerical simulations illustrated the instability caused by capacity drop. We also verified the new model through an empirical study.

This new model of capacity drop is devoid of the fallacy of models based on discontinuous fundamental diagrams. In addition, this study clarifies the causal relationship between capacity drop and ``discontinuous" fundamental diagrams. First, ``discontinuous'' fundamental diagrams are actually incomplete observations of flow-density relations. Second, capacity drop is not an effect of ``discontinuous'' fundamental diagram, but a cause. Third, lane drop is not the only cause of such incomplete observations of fundamental diagrams, as lane-drop, merging, and other bottlenecks can also lead to ``discontinuous'' fundamental diagrams. We have to compare the stable discharging flow-rates before and after the formation of upstream queues as in \reff{fig:I405S-Jeffrey} in order to identify capacity drop.
Therefore, ``discontinuous'' fundamental diagrams cannot be used to model or uniquely identify capacity drop. 

The model can be simply extended for a merging bottleneck, whose upstream demands are $d_1(t)$ and $d_2(t)$ and the downstream supply is $s_3(t)$:
\bsq
\bqn
q_3(t)&=&\min\{d_1(t)+d_2(t), \tilde s_3(t)\},\\
q_1(t)&=&\min\{d_1(t),\max\{\tilde s_3(t)-d_2(t),\alpha \tilde s_3(t)\},\\
q_2(t)&=&\min\{d_2(t),\max\{\tilde s_3(t)-d_1(t),(1-\alpha) \tilde s_3(t)\},
\eqn
\esq
where $\alpha$ is the merging priority for upstream link 1, and the modified downstream supply $\tilde s_3(t)=\min\{s_3(t), C_3(1-\Delta\cdot I_{d_1(t)+d_2(t)>s_3(t)})\}$ with the downstream link's capacity $C_3$. This model extends the priority-based merge model \citep{daganzo1995ctm,jin2010merge}, but the downstream link's capacity drops to $C_3(1-\Delta)$ when the sum of upstream demands is greater than the downstream supply. Note that the capacity reduction effect of lane-changing activities can be captured in the downstream supply $s_3(t)$  \citep{jin2013multi}.

However, this model does not capture the mechanism of capacity drop and is phenomenological since (i) the magnitude of $\Delta$ is exogenous and has to be calibrated for each study site; (ii) capacity drop occurs at one point, as shown in \reff{lane-drop}, but a transition region of 1-2km long can usually be observed around an active bottleneck with capacity drop  \citep{cassidy1999bottlenecks,cassidy2005merge}; and (iii) capacity drop occurs immediately after the upstream is congested, but in reality only after a number of vehicles queue up on the shoulder lane and lane changes disrupt traffic on all lanes \citep{cassidy2005merge}. 
Furthermore, we will be interested in analyzing traffic dynamics inside the transition region during the transition period. We will also be interested in studying the impacts of drivers' accelerating, lane-changing, and merging behaviors in a merging area.

The new model of capacity drop can be readily incorporated into the Cell Transmission Model \citep{daganzo1995ctm,lebacque1996godunov} and used to simulate impacts of capacity drop on the overall traffic dynamics in a road network. 
Therefore it can be used to analyze and simulate how congestion evolves on a road network when a number of bottlenecks interact with each other and how stochastic demand patterns can induce traffic breakdown at various active bottlenecks. Therefore, the new model can be used to evaluate and develop traffic control strategies, including variable speed limits and ramp metering, to delay or avoid the occurrence of capacity drop. 
For example, in \citep{jin2013_vsl}, the new capacity drop model in \refe{cd-entropy} was incorporated into the LWR model and the link queue model \citep{jin2012_link} to design variable speed limits strategies, which were shown to substantially  mitigate traffic congestion at lane-drop bottlenecks.

\section*{Appendix A. Proof of Theorem \ref{thm:cd}}
{\em Proof}. From the feasibility conditions on stationary and interior states, we can see that $q\leq d_1$ and $q\leq s_2$. Therefore, $q\leq\min\{d_1,s_2\}$. We first solve the flow-rate in the following four cases.
\ben
\item When $d_1\leq \min\{s_2, C_*\}$, we assume that $q<d_1$. Thus we have $U_1^*=U_1^0=(C_1,q)$ and $U_2^*=U_2^0=(q,C_2)$. Thus $d_1^0=C_1>s_2^0=C_2$. However, from \refe{lwr:e2} we have that $q=C_*$, which contradicts $q_2<d_1\leq C_*$. Thus in this case $q=d_1$.
\item When $C_*< d_1\leq s_2\leq C_2<C_1$, we consider the following three scenarios:
\bi
\item First, if $q=d_1\leq s_2<C_1$, we have $U_1^*=U_1^0=(q,C_1)$. If $d_1<s_2$, then $U_2^*=U_2^0=(q,C_2)$; if $d_1=s_2$, then $U_2^*=(C_2,q)$, and $U_2^0$ is between $(C_2,q)$ and $(q,C_2)$. In this case $d_1^0\leq s_2^0$, which satisfy \refe{lwr:e2}. Thus $q=d_1$, $U_1^*=(q,C_1)$, and $U_2^*=(q,C_2)$ ($d_1<s_2$) or $U_2^*=(C_2,q)$ ($d_1=s_2$) satisfy \refe{lwr:e2}. 
\item Second, if $q< d_1\leq s_2$ and $q\neq C_*$, we have $U_1^*=U_1^0=(C_1,q)$ and $U_2^*=U_2^0=(q,C_2)$, which lead to $d_1^0=C_1>s_2^0=C_2$. However from \refe{lwr:e2} we have $q=C_*$, which contradicts $q\neq C_*$. Thus it is impossible to have that $q<d_1$ and $q\neq C_*$. 
\item Third, if $q=C_*<d_1\leq s_2$, we have $U_1^*=U_1^0=(C_1,q)$ and $U_2^*=U_2^0=(q,C_2)$. Thus $d_1^0=C_1>s_2^0=C_2$, which satisfies \refe{lwr:e2}. Thus $q=C_*$, $U_1^*=(q, C_1)$ and $U_2^*=(q,C_2)$ satisfy \refe{lwr:e2}. 
\ei
Therefore, both $q=d_1$ and $q=C_*$ satisfy \refe{lwr:e2}. However, from \refe{optimal-entropy}, the unique solution of the boundary flux is $q=d_1>C_*$.
\item When $d_1>s_2$ and $s_2\leq C_*$, if $q<s_2$, then $U_1^*=U_1^0=(C_1,q)$ and $U_2^*=U_2^0=(q,C_2)$, which lead to $d_1^0=C_1>s_2^0=C_2$. However from \refe{lwr:e2} we have $q=C_*$, which contradicts $q<s_2\leq C_*$. Thus $q=s_2$.
\item 
When $d_1>s_2>C_*$, we consider the following three scenarios:
\begin{itemize}
  \item First, if $q>C_*$ and $q< s_2<d_1$. Then $U_1^*=U_1^0=(C_1,q)$, and $U_2^*=U_2^0=(q,C_2)$, which lead to $d_1^0=C_1>s_2^0=C_2$. However from \refe{lwr:e2} we have $q=C_*$, which contradicts $q>C_*$.
  \item Second, if $q>C_*$ and $q=s_2<d_1$. Then $U_1^*=U_1^0=(C_1,q)$, $U_2^*=(C_2,q)$, and $d_2^0>q$. Since  $d_1^0=C_1>C_2\geq s_2^0$, from \refe{lwr:e2} we have $q=\min\{s_2^0, C_*\} \leq C_*$, which contradicts $q>C_*$.
  \item Third, if $q<C_*<s_2<d_1$. Then $U_1^*=U_1^0=(C_1,q)$, and $U_2^*=U_2^0=(q,C_2)$, which lead to $d_1^0=C_1>s_2^0=C_2$. However from \refe{lwr:e2} we have $q=C_*$, which contradicts $q<C_*$.
\end{itemize}
Therefore, $q=C_*$.
\een
In all of the four cases, the boundary flux is uniquely solved by
\bqs
q&=&\cas{{ll} d_1, &d_1\leq s_2\\\min\{s_2,C_*\}, & d_1>s_2}
\eqs
Note that \refe{lwr:e2} cannot be used to pick out a unique solution in $q$ when $C_*< d_1\leq s_2$. Therefore, \refe{lwr:e2} is a necessary condition, but not sufficient. In contrast, \refe{optimal-entropy} is both necessary and sufficient. 

From the feasibility conditions on the stationary states,  $U_1^*=(d_1,C_1)$ when $q=d_1$, and $U_1^*=(C_1,q)$ otherwise. Similarly, $U_2^*=(C_2,s_2)$ when $q=s_2$, and $U_2^*=(q,C_2)$ otherwise.\footnote{Note that interior states may not be uniquely solved, but they do not impact the kinematic wave solutions.} That is, the stationary states are uniquely solved. With the stationary states, we can solve the traditional LWR model to find shock or rarefaction waves on each link. 
\eop

\bibliographystyle{elsarticle-harv}


\end{document}